\documentclass[prb,twocolumn, showpacs,superscriptaddress,graphicx]{revtex4-2}
\usepackage{amssymb}
\usepackage{graphicx}
\usepackage{amsmath}
\usepackage{amsfonts}
\usepackage{mathrsfs}
\usepackage{epsfig}
\usepackage{epstopdf}
\usepackage{xcolor}

\usepackage{upgreek}

\usepackage[unicode]{hyperref}
\usepackage{textcomp} %нужен был для °
\usepackage{multirow}
\usepackage{outline}
\usepackage{pmgraph}
\usepackage{float}

\usepackage[english]{babel}

\graphicspath{{ffunc/}{figures/}}
\begin{document}

%\title{Magnetically memorable superconducting resonators}
\title{Magnetically memorable inductance in superconducting multilayer resonators}

\author{R. Tyumenev}
\affiliation{Moscow Institute of Physics and Technology, 141700 Dolgoprudny, Russia}

\author{D. S. Kalashnikov}
\affiliation{Moscow Institute of Physics and Technology, 141700 Dolgoprudny, Russia}

\author{B. V. Fradkin}
\affiliation{Moscow Institute of Physics and Technology, 141700 Dolgoprudny, Russia}
\affiliation{All-Russian Research Institute of Automatics n.a. N.L. Dukhov (VNIIA), 127030, Moscow, Russia}

\author{A. A. Neilo}
\affiliation{Skobeltsyn Institute of Nuclear Physics, Lomonosov Moscow State University, Moscow 119991, Russian Federation}

\author{A.~G.~Shishkin}
\affiliation{Moscow Institute of Physics and Technology, 141700 Dolgoprudny, Russia}
\affiliation{All-Russian Research Institute of Automatics n.a. N.L. Dukhov (VNIIA), 127030, Moscow, Russia}

\author{N.~V.~Klenov}
\affiliation{Moscow Technical University of Communications and Informatics (MTUCI), 111024 Moscow, Russia}

\author{I.~I.~Soloviev}
\affiliation{Skobeltsyn Institute of Nuclear Physics, Lomonosov Moscow State University,
Moscow 119991, Russian Federation}
\affiliation{All-Russian Research Institute of Automatics n.a. N.L. Dukhov (VNIIA),
127030, Moscow, Russia}

\author{A.~A.~Golubov}
\affiliation{Moscow Institute of Physics and Technology, 141700 Dolgoprudny, Russia}
\affiliation{HSE University, 101000 Moscow, Russia}

\author{M.~Yu.~Kupriyanov}
\affiliation{Skobeltsyn Institute of Nuclear Physics, Lomonosov Moscow State University,
Moscow 119991, Russian Federation}
\affiliation{Moscow Institute of Physics and Technology, 141700 Dolgoprudny, Russia}

\author{I. A. Golovchanskiy}
\altaffiliation[Present address: ]{Institute of Physics, Ecole Polytechnique Federale de Lausanne, 1015 Lausanne, Switzerland}
\affiliation{Moscow Institute of Physics and Technology, 141700 Dolgoprudny, Russia}

\author{V.~S.~Stolyarov}
\affiliation{Moscow Institute of Physics and Technology, 141700 Dolgoprudny, Russia}
\affiliation{All-Russian Research Institute of Automatics n.a. N.L. Dukhov (VNIIA),
127030, Moscow, Russia}

\author{A.~S.~Sidorenko}
\affiliation{Technical university of Moldova, Institute of Electronic Engineering and Nanotechnologies MD2028, Chisinau, Moldova}

\author{S.~V.~Bakurskiy}
\email{r4zz@mail.ru}
\affiliation{Skobeltsyn Institute of Nuclear Physics, Lomonosov Moscow State University, Moscow 119991, Russian Federation}

\date{\today }

\begin{abstract}
{Superconductor–ferromagnet hybrid structures with tunable kinetic inductance are promising elements for neuromorphic and quantum computing circuits. We report the fabrication and microwave characterization of split-ring resonators based on Nb/Co/Nb/Co/Nb/Al spin-trigger multilayers and demonstrate a non-volatile spin-valve effect on their resonant properties. Reversal of the relative magnetization orientation of the cobalt layers produces a reproducible shift of the resonant frequency up to 4 MHz at zero applied magnetic field, corresponding to a change in the kinetic inductance of the structure. The incorporation of a proximitized aluminum overlayer is shown to enhance the inductance contrast between the parallel and antiparallel magnetic states by a factor of approximately three relative to structures without this layer. The experimental results are in quantitative agreement with a microscopic model based on the Usadel equations. The demonstrated magnetic memory of the resonant frequency at zero field establishes spin-trigger multilayers as viable field-programmable inductive elements for superconducting digital and neuromorphic circuits.}
\end{abstract}


\maketitle

\section{Introduction}

{Superconducting electronics underpins the development of energy-efficient and 
high-speed components for information and telecommunication systems, including 
neuromorphic computing devices \cite{ishida2021superconductor, schneider2022supermind, 
islam2023review}, quantum and classical supercomputers \cite{siddiqi2021engineering, 
vozhakov2022state, zikiy2023high, cuthbert2022, pot2023nonvolatile}, as well as 
compatible high-sensitivity detectors \cite{semenov2020effect}. One actively developing 
class of such components is} devices with high kinetic inductance 
\cite{bakurskiy2020controlling, peltonen2018hybrid, hazard2019nanowire, 
kalacheva2024kinemon}.

Elements of kinetic inductance, i.e., conductors in which the energy of the current 
is primarily stored in the kinetic energy of the charge carriers rather than in the 
associated magnetic field energy, offer several significant advantages. 
{They can provide a finite inductance value with considerably smaller 
dimensions compared to geometric inductance elements, enabling direct device 
miniaturization \cite{tolpygo2023progress}. Furthermore, the reduced magnetic field 
contribution results in significantly lower effective mutual inductance compared to 
traditional inductive components, which diminishes parasitic coupling in highly 
integrated circuits \cite{tolpygo2023self}. Finally, the magnitude of kinetic inductance 
directly depends on the charge carrier concentration and can be controlled via 
electrical, field, or magnetic interactions \cite{mazin2005microwave, dominjon2016study, 
zhao2017exploring, levy2021subgap, malnou2021three, parker2022degenerate, 
tan2024operation, khalifa2024kinetic, vissers2015frequency, adamyan2016tunable, 
mahashabde2020fast, luomahaara2014Kinetic, vodolazov2023nonlinear}. 
The field- and current-tunability of kinetic inductance has become the basis for} 
a wide class of devices, including photon detectors \cite{mazin2005microwave, 
dominjon2016study, zhao2017exploring, levy2021subgap}, parametric amplifiers 
\cite{malnou2021three, parker2022degenerate, tan2024operation, khalifa2024kinetic}, 
tunable resonators \cite{vissers2015frequency, adamyan2016tunable, mahashabde2020fast}, 
and other devices \cite{luomahaara2014Kinetic, vodolazov2023nonlinear}.

{At the same time,} the development of linear tunable inductance elements 
{offers significant prospects} for digital electronics circuits 
\cite{ruzhickiy2025programmable} and neuromorphic computing devices 
\cite{soloviev2018adiabatic, schegolev2022tunable}, in which an independent method of 
controlling inductance allows switching the operating modes of devices and tuning weight 
characteristics of synapses and neurons. {A key requirement} for such 
devices is stability of the selected state against the influence of the structure's 
operating currents. This can be achieved through the use of magnetic hybrid structures 
\cite{linder2015superconducting, yu2026electromagnetic}, including spin valves 
\cite{leksin2015superconducting, kamashev2024superconducting, karelina2021scalable, 
karelina2024magnetic}. {In these structures, the relative orientation of 
the magnetization vectors of the ferromagnetic (F) layers determines the critical 
temperature and, consequently, the inductance of the structure. However, a 
{fundamental limitation} of such elements for inductive applications is 
the complete transition into the resistive state for one of the magnetization 
orientations. If an operating point is chosen where both states remain superconducting, 
the spin-valve effect on inductance becomes very weak.}

To address this issue for circuits with longitudinal current flow, it was recently 
proposed to use the so-called spin-trigger effect \cite{neilo2024spin, 
neilo2024josephson}. {This concept involves fabricating} an 
S-F$_1$s$_1$F$_1$-s multilayer structure with a set of layers possessing different 
functionalities{: a thick superconducting S layer serves as a source of 
Cooper pairs; the magnetic F$_1$s$_1$F$_1$ trilayer constitutes a spin valve which, in 
the parallel magnetization state, more effectively suppresses pair correlations; and a 
thin s layer undergoes its own superconducting transition only when the valve is open, 
driven by the proximity effect from the source.} In the work \cite{neilo2025magnetic}, to enhance the effect, it was proposed to modify the structure and use an sN bilayer as the element with tunable inductance, where N is a low-resistivity normal metal layer with a long coherence length $\xi_N$. In this case, the inductance of the proximitized sN 
bilayer {is} significantly lower than the inductance of the rest 
structure. As a result, when the spin valve opens, the current flows primarily into the 
sN bilayer, whereas when the valve is closed, the current flows through the source S 
layer, leading to a substantial increase in the total inductance of the structure. 
According to the predictions of \cite{neilo2025magnetic}, such a modification was 
expected to increase the efficiency of the spin-valve effect in terms of inductance change 
by a factor of 3{--}4 compared to a conventional spin-trigger multilayer 
structure without an N layer.

In this work, we investigated the inductive properties of S-F$_1$s$_1$F$_1$-s and 
S-F$_1$s$_1$F$_1$-sN structures by studying the microwave characteristics of split-ring 
resonators (SRRs) based on them, {operating in the 4--7~GHz frequency 
range relevant for superconducting digital, neuromorphic, and quantum circuit 
applications}. We demonstrated the spin-valve effect occurring in these structures upon 
the magnetization reversal of the ferromagnetic layers{, and showed 
non-volatile control of the resonant frequency by} {approximately} 4~MHz 
at zero magnetic field {via applied field pulses}.

\begin{figure}[t]
\begin{minipage}[h]{0.99\linewidth}
\center{\includegraphics[width=\linewidth]{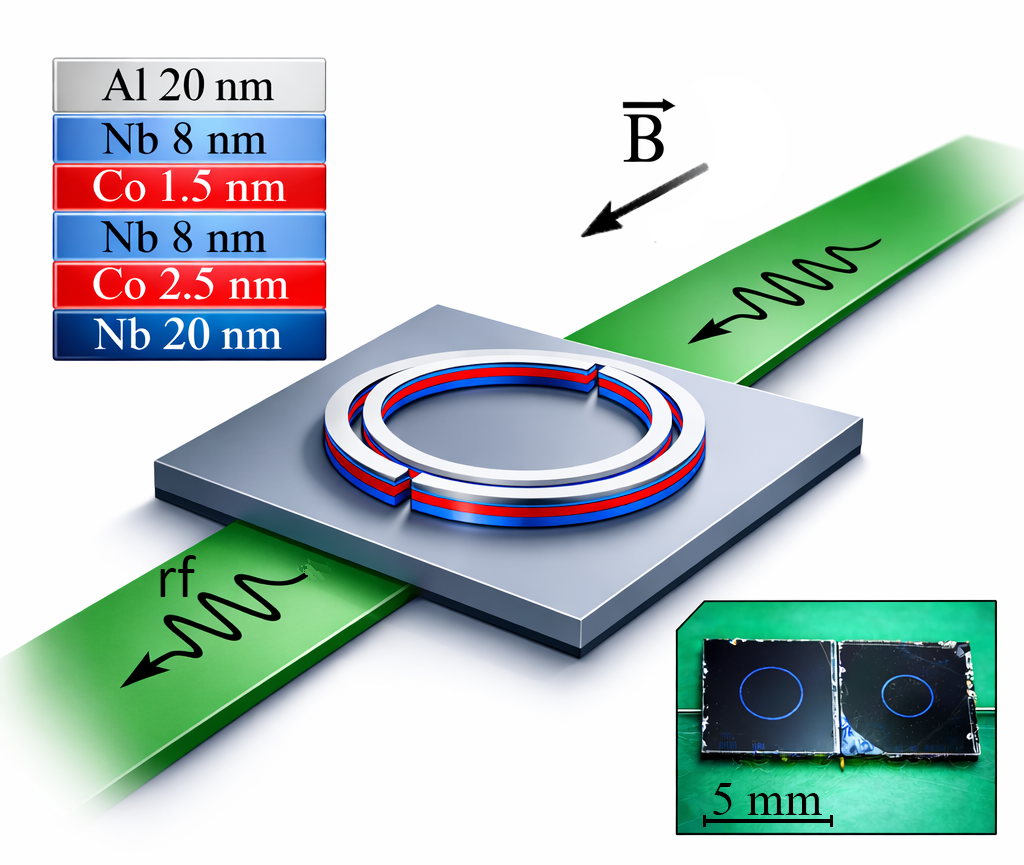}}
\end{minipage}
\caption{Schematic illustration of the studied sample. A multilayer split-ring resonator (SRR) on a Si (100) substrate is placed on top of the central line of a planar waveguide (strip-line type). The external magnetic field $B$ is applied along the waveguide. The top-left inset shows a cross-sectional sketch of the sample with the actual layer thicknesses. The bottom-right inset shows a photograph (top view) of the measured samples.}
\label{figure: scematic}
\end{figure}

\begin{figure*}[t]
\begin{minipage}[h]{0.99\linewidth}
\center{\includegraphics[width=0.99\linewidth]{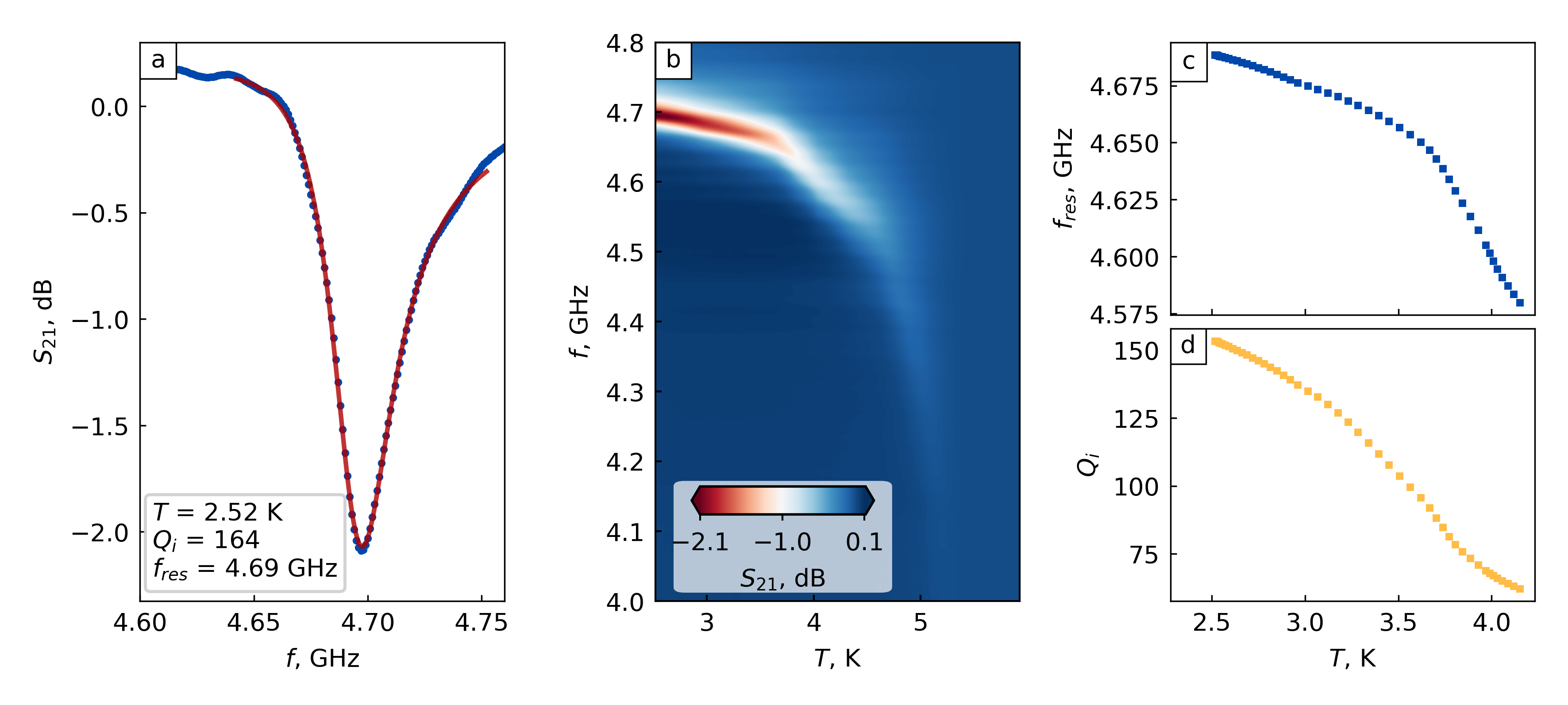}}
\end{minipage}
\caption{Temperature dependences of the parameters of the 4.7~GHz multilayer resonator. (a) Transmission coefficient $S_{21}$ of the resonator at a fixed temperature of $2.5$~K after background subtraction. The red curve shows a fit using the asymmetric Lorentzian profile \eqref{eq: S21}. The resonator parameters obtained from the fit are indicated in the legend. (b) Temperature dependence of $S_{21}(f)$. (c) Resonant frequency and (d) internal quality factor of the resonator as functions of temperature.}
\label{figure:S21(T)}
\end{figure*}

\section{Experiment}

\subsection{Samples and measurement methods}

To study the influence of the relative orientation of the magnetization vectors of the layers on the kinetic inductance of the system, structures based on split-ring resonators (SRRs) were fabricated; a schematic of these resonators is shown in Fig.~\ref{figure: scematic}. This type of resonator consists of two concentric metal rings with gaps located on opposite sides. The resonators were fabricated from a Nb(20)/Co(2.5)/Nb(8)/Co(1.5)/Nb(8)/Al(20) multilayer structure, deposited by magnetron sputtering onto a 0.5 mm thick Si(100) substrate, followed by lift-off {process}. Several series of resonators were fabricated, designed for different resonant frequencies in the range of about 4 and 7 GHz, which allowed them to be investigated independently. The fabrication process is described in more detail in Supplementary Materials S1. Additionally, Nb(20)/Co(2.5)/Nb(8)/Co(1.5)/Nb(8) samples without the top aluminum layer and reference samples consisting of a single 100 nm thick niobium layer were fabricated and measured.

During measurements, the samples were placed on a microstrip {feedline}, as shown in Fig. \ref{figure: scematic}. {Measurements were performed in a longitudinal magnetic field generated by a superconducting coil, with the microwave transmission coefficient $S_{21}(f)$}.

\subsection{Measurement of resonant absorption}

{Fig.~\ref{figure:S21(T)}a shows the measured $S_{21}(f)$ spectrum 
at $T = 2.5$~K.} The background, measured separately at a higher temperature of 
$6.3$~K, is subtracted from the raw data. {To extract resonator 
parameters quantitatively, we fit} the experimental data with an asymmetric Lorentzian 
profile \cite{fit_res}:

\begin{equation}
    \label{eq: S21}
    S_{21} = 1 - \frac{(Q_l / |Q_c|)e^{i\phi}}{1 + 2i Q_l \left({f}/{f_{res}} - 1 
    \right)}.
\end{equation}
{Here $f$ is the probe frequency, $f_{res}$ is the resonance frequency, $Q_l$ is the loaded quality factor accounting for both internal losses
and coupling to the feedline,  $Q_i$ is the internal quality factor, $Q_c$ is the complex coupling quality factor
characterising the energy leakage rate into the transmission line,
and $\phi$ is a correction angle arising from the impedance mismatch between the resonator and the feedline 
\cite{Khalil_2012}.}

The SRR resonator was designed to have a resonant frequency of 5~GHz, considering 
only geometric inductance. Consistent with this design, a pure niobium sample shows 
a frequency close to this value (see Suppl. Mat. and Fig. S1). In 
contrast, the multilayer SFsFsN structure exhibits a much lower resonant frequency 
of $f_{res} \approx 4.7$~GHz {(the fit of of experimental $S_{21}(f)$ dependence  by  the asymmetric Lorentzian profile \eqref{eq: S21} is shown by the red curve in 
Fig.~\ref{figure:S21(T)}a)}. {This deviation is attributed to the 
enhanced kinetic inductance of the thin multilayer structure due to the ferromagnetic 
proximity effect.} {Estimated from the frequency shift relative to 
the reference Nb resonator, the kinetic inductance of the multilayer structure is 
$L_k \sim 1$~nH, compared to a total inductance of $L \sim 8$~nH, which allows 
observable effects associated with changes in the kinetic contribution.} The quality factor of the resonator made from the multilayer structure is an order of magnitude lower than that of the niobium resonator (compare Fig.~\ref{figure:S21(T)}d and Suppl.Mat Fig.S2d). {The reduced $Q_i$ is likely 
attributable to} inhomogeneities in the thicknesses of the structure's layers, as 
well as the domain structure of the ferromagnetic films, which leads to 
{an inhomogeneous current distribution} and broadening of the 
resonant peak. Another possible reason for the reduced quality factor is the 
suppressed gap and the formation of subgap subbands in the density of electronic 
states of the ferromagnetic layers \cite{seleznyov2024intrinsic}, which leads to 
quasiparticle excitation {by the microwave probe signal} and 
subsequent quasiparticle poisoning of the resonator.

With increasing temperature, the resonant peak shifts to lower frequencies and 
decreases in amplitude, as can be seen in Fig.~\ref{figure:S21(T)}b. The signal 
completely {vanishes} at a temperature of approximately 5.55~K, 
{marking the superconducting transition at $T_c \approx 5.55$~K.} 
The critical temperature is reduced below the standard bulk niobium value of 9.2~K 
{due to 
the pair-breaking inverse proximity effect from the adjacent ferromagnetic films.} 
{Fitting each spectrum with Eq.~\eqref{eq: S21}, we extract the 
temperature dependencies of $f_{res}$ and $Q_i$, shown in 
Fig.~\ref{figure:S21(T)}c,d.} Both quantities decrease monotonically with increasing 
temperature. {To maximize the 
resonator sensitivity to magnetic-state changes,} we performed further measurements 
in a magnetic field at the lowest possible temperature of $T = 2.5$~K for this 
experiment.

\begin{figure*}[t]
\begin{minipage}[h]{0.99\linewidth}
\center{\includegraphics[width=0.99\linewidth]{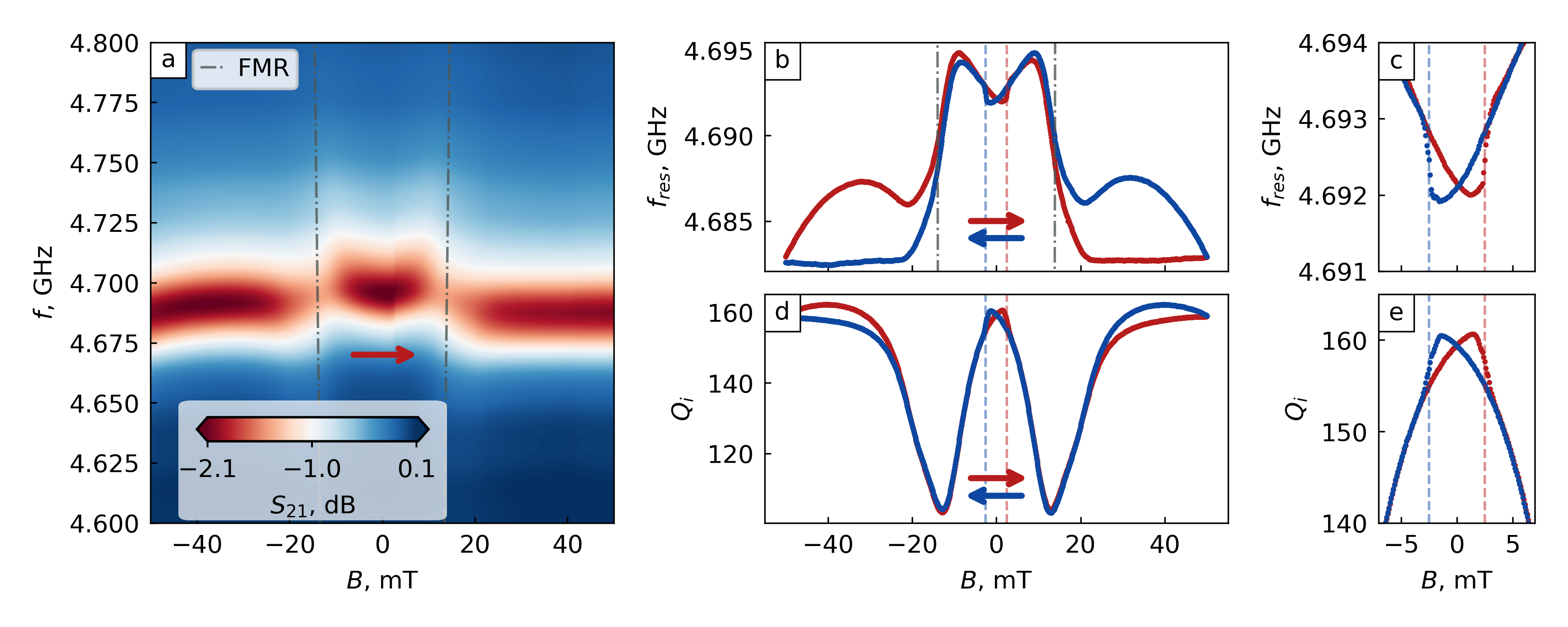}}
\end{minipage}
\caption{Dependence of the parameters of the 4.7~GHz multilayer resonator on the external magnetic field. (a) Measured transmission coefficient $S_{21}$ as a function of the magnetic field after background subtraction. The dotted-dashed line indicates the FMR dependence. (b) Resonant frequency and (d) internal quality factor, extracted using the fit from Eq.~\eqref{eq: S21}, as functions of the external magnetic field. The dashed line marks the field at which the thinner ferromagnetic layer reverses, resulting in an antiparallel configuration. (c) Resonant frequency and (e) internal quality factor in an expanded view of the region corresponding to the reversal of the thin layer.}
\label{figure:magnetic_4,7}
\end{figure*}

\subsection{Measurements in a magnetic field}

To investigate the effect of the ferromagnetic layers on the {resonant 
properties and kinetic inductance of the} system, we performed a series of measurements 
in a longitudinal magnetic field. The resulting $S_{21}(B)$ dependence is shown in 
Fig.~\ref{figure:magnetic_4,7}(a), with the field swept from negative to positive 
values. When the field sweep direction was reversed, we obtained a pattern symmetric 
with respect to $B = 0$. {Two features are visible in the data. The 
first is} the suppression of the resonant peak amplitude and the shift of the resonant 
frequency in fields around $\pm 15$~mT. {This behavior is presumably 
attributed to} the anticrossing of the resonator {mode} and 
ferromagnetic resonance (FMR) in the cobalt films. {Similar measurements 
on an SRR} made from the same multilayer film with a designed resonant frequency of 
7~GHz {revealed anticrossing features at higher fields of 
$\pm 20$~mT} (see Suppl. Mat. Fig.S2). The FMR frequency in samples with 
magnetization vector $M_{eff}$ and external magnetic field $B$ lying in the plane of the film can be determined by the Kittel formula \cite{Kittel_1948}: 
$f_{\text{FMR}} = \gamma \sqrt{B(B+\mu_0 M_{eff})}$. The {black vertical} dashed lines in 
Fig.~\ref{figure:magnetic_4,7} show the dependence $f_{\text{FMR}}(B)$, 
{calculated using $\gamma = 28$~GHz/T (the free-electron value). The 
value $\mu_0 M_{eff} = 1.8$~T was determined by fitting the anticrossing positions 
simultaneously} for the two SFsFsN samples with resonant frequencies 
$f_{res} \approx 4.7$~GHz and $f_{res} \approx 6.5$~GHz. The obtained numerical value 
of the magnetization is consistent with the standard value for cobalt thin films 
\cite{Beaujour_2006}. {This confirms the anticrossing assignment of the 
features} in Fig.~\ref{figure:magnetic_4,7}(a).

The second feature in Fig.~\ref{figure:magnetic_4,7}a {is} a sharp 
jump in the resonance frequency at a positive field of $\approx 2.5$~mT. 
{To quantitatively} describe our results, we {fitted} 
the resonant peaks {using Eq.~\eqref{eq: S21}} for each value of the 
magnetic field and obtained {the dependences $f_{res}(B)$ and $Q_i(B)$}, 
which are shown in Fig.~\ref{figure:magnetic_4,7}(b) and (d), respectively. 
Panels~\ref{figure:magnetic_4,7}(c) and (e) show the same dependences in a narrow 
range of fields. {The jump in $f_{res}(B)$ is symmetric,} occurring at 
$B = 2.5$~mT when sweeping from negative to positive values and at $B = -2.5$~mT when 
sweeping in the opposite direction. {Both the switching field and the hysteretic behavior are consistent with the magnetization reversal of the thinner cobalt layer.} At these same fields, the quality factor stops increasing and 
{drops sharply.} Moreover, the curves plotted in 
Fig.~\ref{figure:magnetic_4,7}(b) and (d) for opposite field sweep directions 
{reveal} global hysteresis in the system {up to 
$\pm 50$~mT}. We attribute this behavior to the influence of magnetic domains in the 
thick cobalt layer, which we discuss in detail in the next section.

\subsection{Memory effect}

\begin{figure}[t]
\begin{minipage}[h]{0.99\linewidth}
\center{\includegraphics[width=0.99\linewidth]{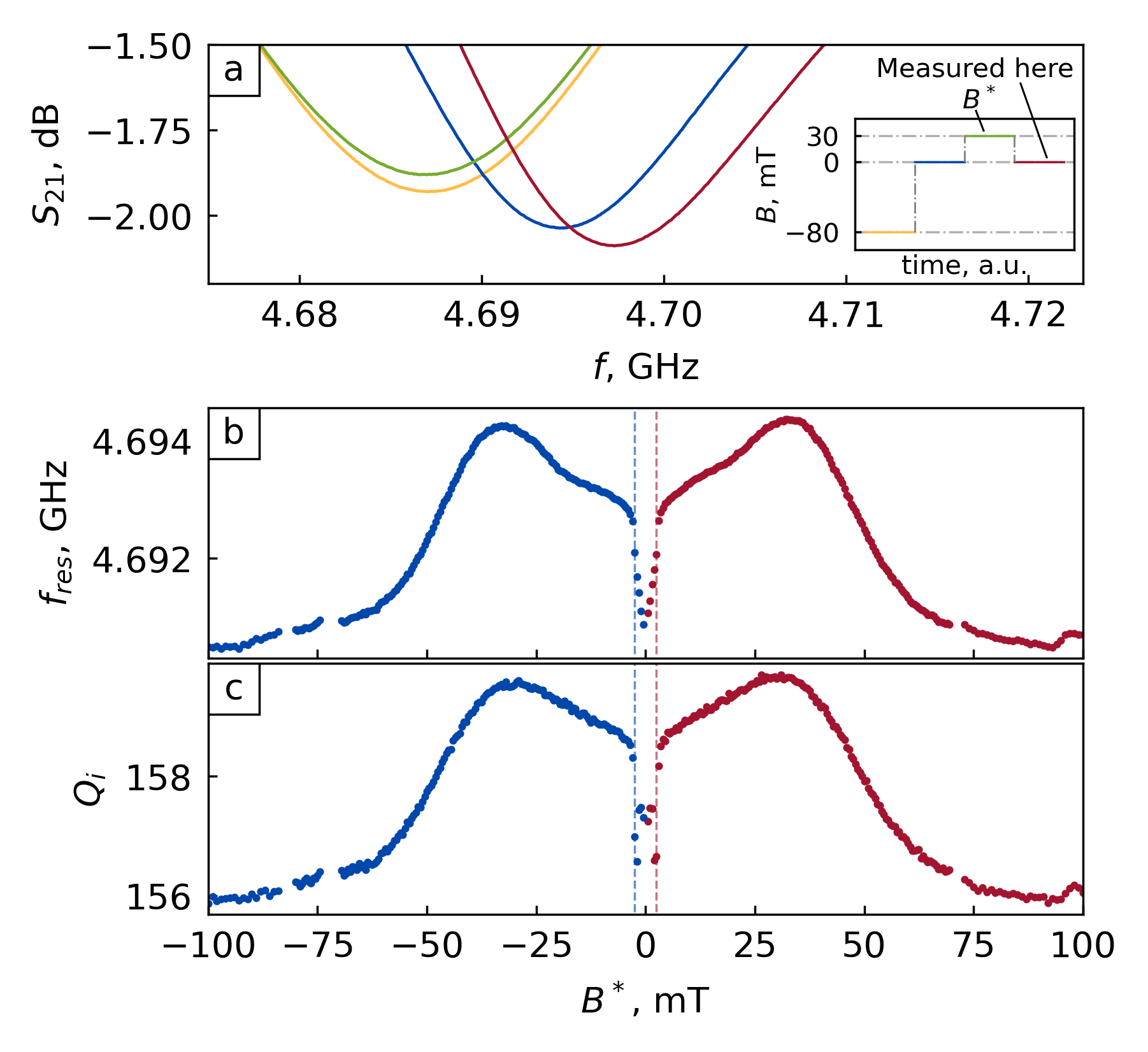}}
\end{minipage}
\caption{ Characteristics of the 4.7~GHz multilayer resonator depending on the history of the applied external magnetic field. (a) Transmission coefficient $S_{21}$ under a field $B_1 = -80$~mT (yellow curve); after switching the field off, the sample is in the P-state (blue curve); under a field $B = 30$~mT (green curve); and after switching the field off, the sample is in the AP-state (red curve). The inset shows the history of the applied external magnetic field. Dependence of the resonant frequency (b) and the internal quality factor (c) of the multilayer resonator, measured at zero field after full magnetization reversal and the temporary application of a finite field $B^*$ (indicated on the x-axis). Blue symbols show the measurement results for full magnetization reversal in a +80~mT field followed by a temporarily applied field in the negative direction. Red symbols show the dependence for full magnetization reversal in a negative field followed by a temporarily applied field in the positive direction. The dashed line indicates the field $\pm$2.5~mT at which the reversal of the thinner ferromagnetic layer occurs.}
\label{figure:memory_effect}
\end{figure}

The irreversible behavior of the resonator characteristics {allows this system to function as a two-state inductive 
memory element.} {To characterize this memory effect,} we conducted a series of experiments in which the {characteristics} of resonator was measured at zero magnetic field after {controlled reorientation of the magnetization in the cobalt layers.} 
An example {is shown} in Fig.~\ref{figure:memory_effect}~a, which shows the transmission coefficient $S_{21}$ in the vicinity of the resonant frequency. 
Initially, a field $B_1 = -80$~mT was applied to the sample, sufficient to magnetize both cobalt layers negatively. We then measured the resonance curve {at $B = 0$}, shown by the blue line in Fig.~\ref{figure:memory_effect}~a. In this state, 
both layers are magnetized parallel. To achieve an antiparallel orientation of the {magnetic } moments, we applied a positive field of 30~mT and then measured the  {$S_{21}$} again at 
zero field (red curve). {In the antiparallel state, $f_{res}$ is 
4~MHz higher than in the parallel state.} From the measurements shown in 
Fig.~\ref{figure:magnetic_4,7}, it is evident that the antiparallel magnetization 
state already occurs at a field of 2.5~mT. However, {the field of 
30~mT was chosen because at this value the magnetization of the thicker layer does not 
yet reverse,} and the effect becomes more noticeable, as will be shown below.

More detailed measurements of the memory effect were performed as a function of the 
{magnetizing} field {$B^*$} {following the protocol 
described below.} First, a large negative longitudinal magnetic field, $B = -100$~mT, 
was applied to ensure {full} alignment of both magnetic layers in the 
negative (parallel) direction. The field was then switched off, and the resonant 
frequency was measured at $B = 0$ in this parallel state. Subsequently, a finite 
positive longitudinal field $B^*$, varied in the range from 0 to 100~mT, was applied. 
After removing this field, the resonant response was measured again at zero field in 
the resulting magnetic configuration. {The same protocol} was carried 
out for negative values of $B^*$, starting from an initial state prepared in a large 
positive field $B = 100$~mT.

Figures~\ref{figure:memory_effect}b,c show the dependence of the resonant frequency 
and quality factor of the multilayer structure on the magnetic-field history. The 
resonant frequency exhibits a clearly non-monotonic dependence on $B^*$. At large 
fields, $B^* > 80$~mT, the frequency saturates at a constant value corresponding to 
the parallel state{, regardless of the sign of the initial 
magnetization.} For small positive fields, the resonant frequency increases rapidly, 
exhibiting a discrete jump of approximately 0.5~MHz at $B^* \approx 2.5$~mT. This 
feature, similar to those observed in Fig.~\ref{figure:magnetic_4,7}(c,e), is 
attributed to the magnetization reversal of the thin cobalt layer. With further 
increase of the applied field, the frequency continues to rise over a broad range up 
to $\sim 30$~mT, reaching a total shift of {$\sim 4$~MHz.} We 
attribute this gradual increase to the formation of magnetic domain states in one of 
the ferromagnetic layers. {Near domain walls, the effective exchange 
field is spatially averaged over the superconducting coherence length in the aluminum 
layer. This reduces the pair-breaking influence of the ferromagnet,} enabling the 
formation of additional low-inductance current channels and thereby increasing the 
resonant frequency. This interpretation is further supported by the observation that 
the frequency tuning in this field range {builds up} progressively 
over multiple magnetization reversal cycles, rather than occurring in a single step. 
Moreover, similar domain-state behaviour has been reported 
previously~\cite{domain_Krasnov}.

\begin{figure}[t]
\begin{minipage}[h]{0.99\linewidth}
\center{\includegraphics[width=0.99\linewidth]{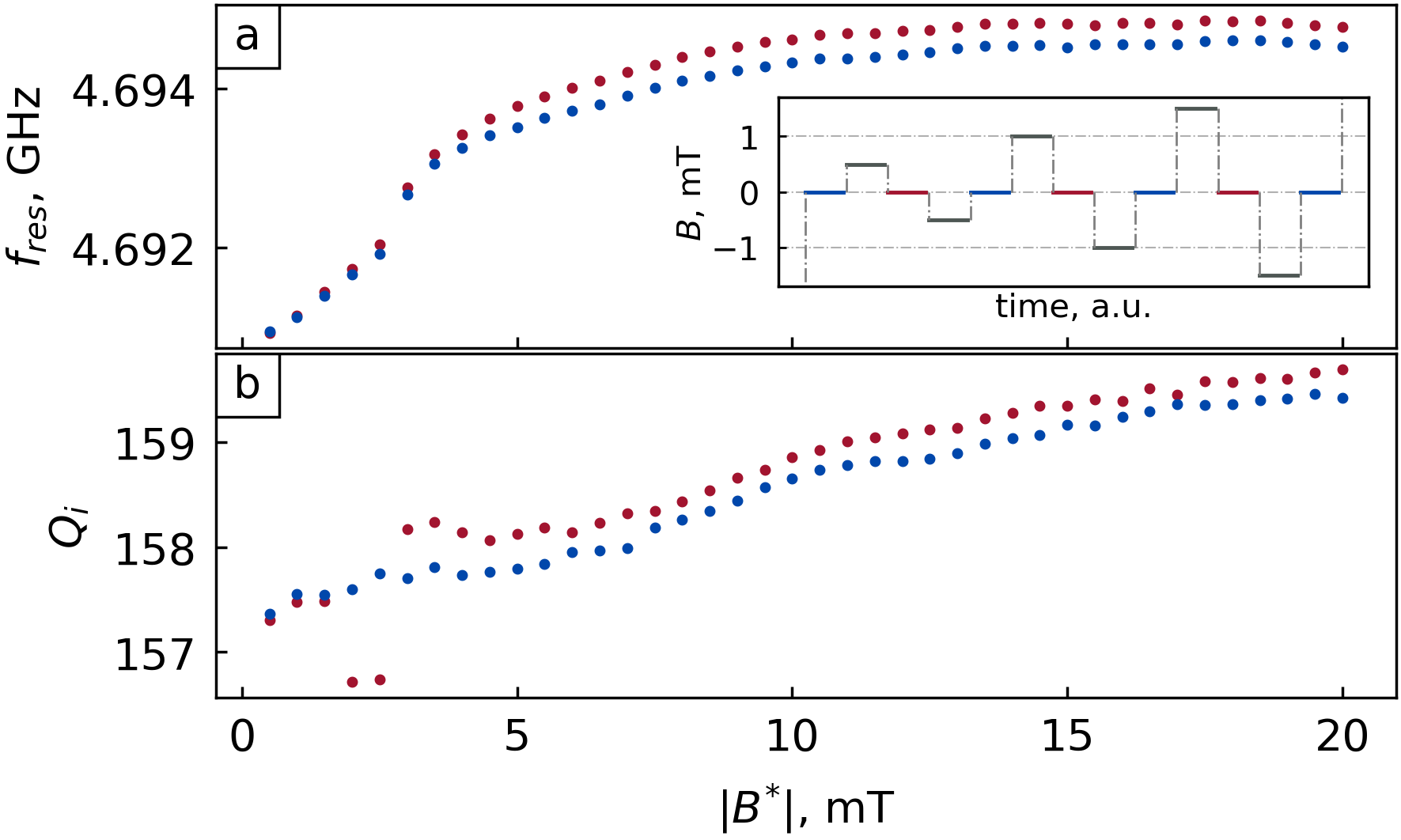}}
\caption{Dependence of the resonant frequency (a) and the internal quality factor (b) of the multilayer resonator, measured at zero field after full magnetization reversal and the temporary application of a finite field $B$, whose value is indicated within the panel. Blue (red) points show the measurement results at zero field after the temporary application of a field in the negative (positive) direction. The inset illustrates the history of the applied external magnetic field. After applying a field of 2.5 mT, at which the reversal of the thinner ferromagnetic layer occurs, a sharp jump in the resonant frequency change of approximately 0.7 MHz is observed for both the red and blue points. Notably, applying a negative pulse does not return the system to its initial state. }
\label{figure:both_hops}

\end{minipage}
\end{figure}

{To verify the domain-state interpretation,} we performed another type of memory effect measurement 
with sequential switching of the thin layer using increasing magnetic field pulses of 
alternating polarity{, following the protocol described below.} First, 
the system was initialized into a parallel magnetic state by applying a field of 
$B_1 = -80$~mT. Next, magnetic field pulses of alternating polarity {with 4~s duration} were applied 
sequentially. After each pulse, the field was turned off, and the resonant frequency 
and quality factor were read out at zero field. The amplitude of each subsequent pulse 
was increased by 0.5~mT. The results {are shown} in 
Fig.~\ref{figure:both_hops}. {Fields below 1.5~mT have negligible 
effect on} the resonant frequency and do not lead to a noticeable memory effect. 
{Pulses exceeding $\sim 5$~mT} lead to an increase in the resonant frequency by {$\sim 4$~MHz}, and applying a negative pulse does not return the system to its initial state{, indicating that the domain configuration is not simply reversed by a single field pulse.} The measurement results for the resonant frequency after applying a negative field (blue points) and a positive 
field (red points) show a similar dependence $f_{res}(B^*)${; the 
$\sim 0.5$~MHz offset between the two curves is attributed to the contribution of the  thin-layer magnetization reversal, while} the contribution from the formation of the  domain structure is larger, amounting to {$\sim 4$~MHz.} This contribution corresponds to the formation of domain states in one of the cobalt layers and requires a series of pulses with amplitudes of 
{20--40~mT to be established, depending on the pulse amplitude rather than polarity.}

\begin{figure}[h!]
\begin{minipage}[h]{0.99\linewidth}
\center{\includegraphics[width=0.99\linewidth]{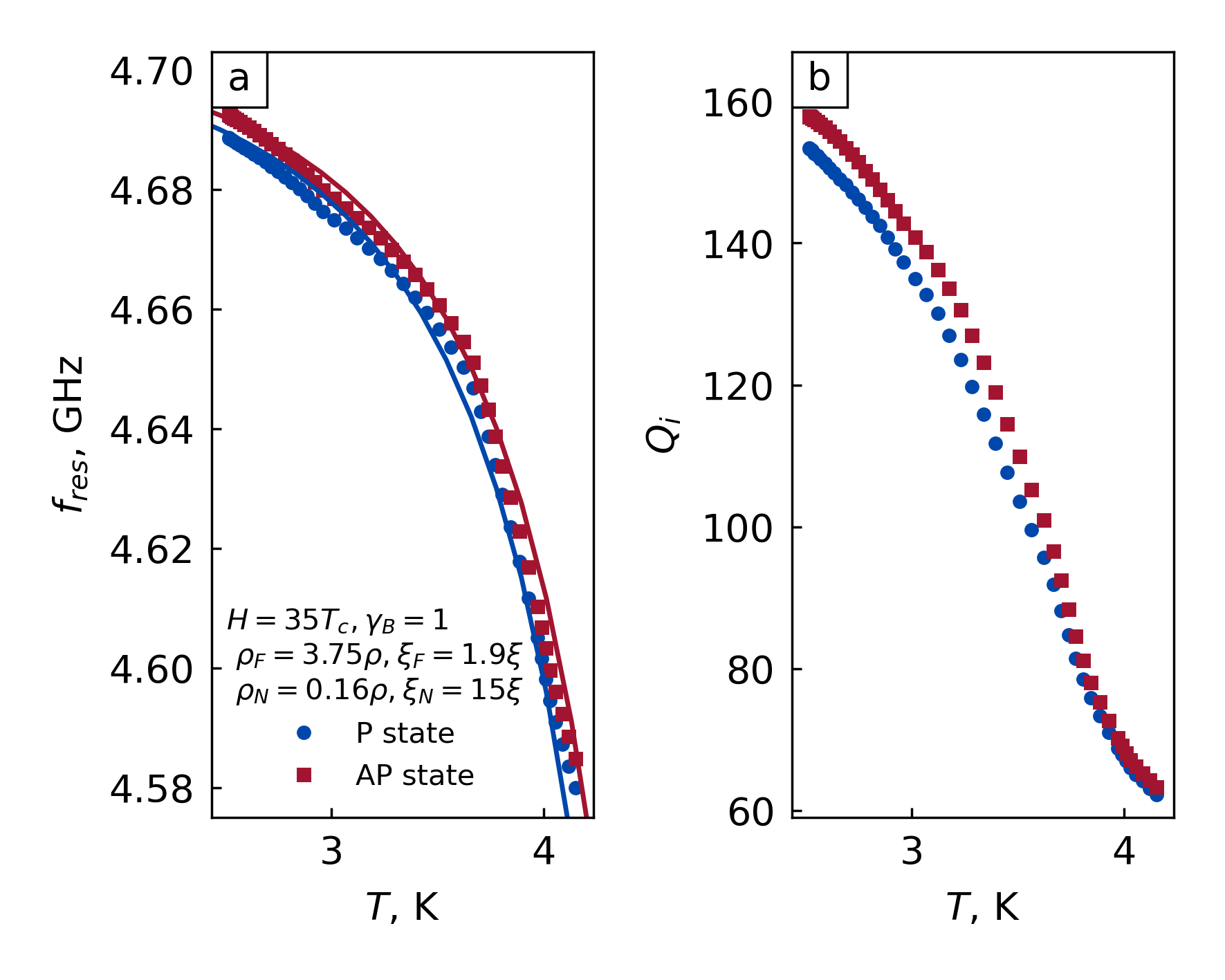}}
\end{minipage}
\caption{Temperature dependence of the resonant frequency (a) and the internal quality factor (b) of the resonator based on the Nb(20)-Co(2.5)-Nb(8)-Co(1.5)-Nb(8)-Al(20) multilayer structure, measured in the state with parallel (blue points) and antiparallel (red points) magnetization orientation. The solid lines represent a fit within the framework of the microscopic model using the parameters listed in the figure legend.}
\label{figure:fit_T}
\end{figure}

{The memory effect persists over the entire superconducting temperature range, up to $T_c$} (Fig.~\ref{figure:fit_T}). {Throughout this range,} the AP state is characterized by a higher resonant frequency and quality factor than {the P state.} {These results are described using} a 
microscopic model (see Suppl. Mat.) {by computing 
the spatial profile of the pairing amplitude for each magnetization state, from which 
the sheet inductance and resonant frequency are derived.} The {fit 
(solid lines in Fig.~\ref{figure:fit_T}) yields} the parameters of exchange energy
$H_{Ex}=35 ~ T_C,$,  interface suppression $\gamma_B=1$, resistivity $\rho_F=3.75 ~ \rho_S, $ $\rho_N=0.16 ~ \rho_S,$ and coherence length 
$\xi_F = 1.9 ~ \xi_S,$ $\xi_N = 15 ~ \xi_S $ of the N and F layers, {in good agreement with the parameters extracted from fitting} the dependence of the critical temperature of the Nb($x$)/Co(2.5) wedge sample on the superconductor layer thickness 
(see Suppl. Mat. Fig. S4). 

% \textcolor{violet}{Экспериментальная процедура заключалась в следующем. Исходно структура была переведена в состояние с параллельной намагниченностью путем воздействия сильного продольного магнитного поля, направленного в отрицательном направлении, с последующим его снятием. После этого была измерена резонансная частота в состоянии остаточной намагниченности. Затем к образцу прикладывалось положительное магнитное поле заданной величины, после снятия которого проводились повторные измерения. Далее последовательность повторялась для магнитного поля той же величины, но направленного в отрицательном направлении. На каждом последующем шаге итерации величина положительного магнитного поля увеличивалась с шагом 0.5 мТ, и описанная выше процедура измерений повторялась. Цикл перемагничивания был проведен вплоть до достижения максимального поля величиной 20 мТ. }

\section{Discussion}

{Thus, we have demonstrated} a significant memory effect in the 
Nb(20)/Co(2.5)/Nb(8)/Co(1.5)/Nb(8)/Al(20) structure, {of up to} 
4~MHz {upon full magnetization reversal from the parallel state, and 
$\sim 0.5$~MHz upon} sequential switching of the thin ferromagnetic layer by small 
fields.

\begin{figure}[t]
\begin{minipage}[h]{0.99\linewidth}
\center{\includegraphics[width=0.99\linewidth]{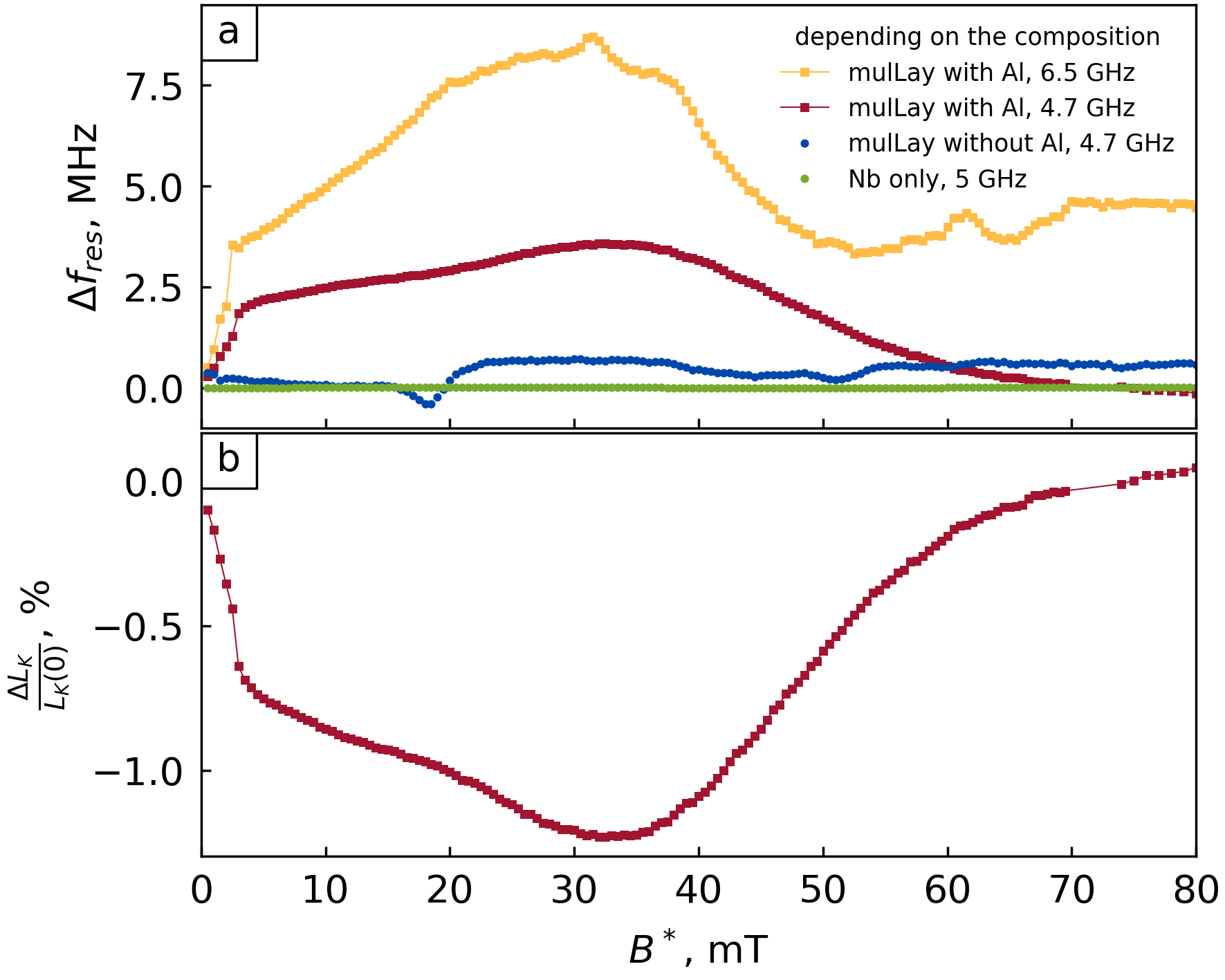}}
\caption{(a) Change in the resonant frequency at zero magnetic field as a function of the temporary magnetization in an external positive field (indicated on the x-axis) for three different layer compositions of the SRR resonator at a fixed temperature of $2.5$~K. (b) Relative change in the kinetic inductance at zero magnetic field as a function of the temporary magnetization in an external positive field (indicated on the x-axis) for the 4.7~GHz multilayer resonator. The geometric inductance and capacitance of the resonator were calculated using EM simulation.}
\label{figure:compositon}
\end{minipage}
\end{figure}

{The physical origin of this memory effect warrants further 
discussion. To this end, we compared} the shift of the resonant frequency depending 
on the field history in the studied sample with control samples: a split-ring resonator 
{fabricated from a single niobium layer} and one {
fabricated from} the multilayer magnetic structure Nb(20)/Co(2.5)/Nb(8)/Co(1.5)/Nb(8) 
{without the top aluminum layer.} For each of the structures, 
measurements were carried out according to the protocol presented above, 
{initializing the P-state with a large negative field and subsequently 
applying positive field pulses,} with the resonant frequency measured after the field 
was turned off. The results are shown in Fig.~\ref{figure:compositon}.

Measurements of the SRR {fabricated from} niobium (green line in 
Fig.~\ref{figure:compositon}) {reveal no noticeable memory effect.} 
{This confirms that the observed effects originate from the magnetic 
subsystem of the structure,} ruling out Abrikosov vortex pinning and flux motion as 
{the primary mechanism.}

{SRRs fabricated from} the multilayer structure without the top Al 
layer (blue points in Fig.~\ref{figure:compositon}) {exhibit a memory 
effect, with a resonant frequency shift not exceeding} approximately 1--1.5~MHz. 
{This is reduced by a factor of $\sim 3$} compared to analogous 
structures incorporating an aluminum layer, in full agreement with the predictions of 
the microscopic model proposed in Ref.~\cite{neilo2025magnetic}. {
Notably, the resonant frequency variation in structures without the aluminum layer is 
stochastic:} the effect is not reproducible from sample to sample and {
exhibits no symmetry under field reversal.} In contrast, for samples with the metallic 
coating, the functional dependence $\Delta f_{\mathrm{res}}(B^*)$ is reproducible 
across different resonators and remains symmetric under a change in the sign of the 
applied field. More detailed measurements of the memory effect in SRRs fabricated from 
multilayer structures without the aluminum interlayer are presented in 
Supplementary Materials.

{These properties are attributed to the non-reproducible domain structure forming in at least one of the ferromagnetic layers}. In samples without the aluminum coating, a 
substantial fraction of the {supercurrent} flows through the magnetic 
FsF subsystem. Consequently, the spatial distribution of the current {
is determined by} the configuration of domain walls in the ferromagnetic layers, which 
is inherently non-reproducible from sample to sample. In contrast, the low-resistivity 
aluminum layer effectively shunts the ferromagnetic structure{, such 
that, when the spin valve is open, the supercurrent flows primarily through the aluminum 
layer.} The presence of a complex domain structure in this case determines only 
{the degree to which the Al layer is proximized by superconducting 
correlations,} which indirectly affects its inductance and ability to carry supercurrent. 
{This spatial averaging averages out the stochastic variations in domain 
structure formation and leads to reproducible results in the memory effect measurements.}

\section{Conclusion}

{We have demonstrated, at the proof-of-principle level, that spin-triggered S-FsF-sN structures can function as magnetic-memory-based tunable inductance elements. Depending on the magnetic prehistory and the resulting magnetic configuration, the resonance frequency of an SRR fabricated from such a hybrid structure can be shifted by several MHz.}

{The inclusion of a metallic shunt layer enhances the effect and, importantly, improves its reproducibility by compensating for the intrinsic non-reproducibility of the ferromagnetic domain structure.}

%Such a tunable inductance element with {in-plane supercurrent} flow can be used to create efficient synapses and neurons, that is, highly integrated components requiring individual tuning. These structures can also be utilized as tunable resonators.

{Such a tunable inductance element, operating with in-plane supercurrent flow, can be used to implement synapses and neurons in highly integrated neuromorphic superconducting circuits, and can also serve as a tunable microwave resonator.}

{
Further improvement of both the spin-valve effect amplitude and the resonator quality factor can be anticipated through targeted circuit-level optimization, including lithographic co-integration of the resonator and feedline on a single chip. In addition, the use of ferromagnetic alloys with a reduced exchange field is expected to suppress subgap quasiparticle excitations and enhance device performance. More broadly, these strategies open a pathway toward scalable, low-loss superconducting circuits with embedded magnetic functionality, enabling adaptive and reconfigurable elements for next-generation neuromorphic and microwave devices.}

%An effective enhancement of the spin-valve effect strength and the quality factor of such resonators is possible through circuit design solutions, for example, by lithographically defining the resonator on the same chip as the feedline, as well as by using ferromagnetic alloys with a smaller exchange field, which opens up broad opportunities for the development of this class of devices.

\section*{Acknowledgements} 

Fabrication and measurements were supported by the Ministry of Science and Higher Education of the Russian Federation (Agreement No. 075-15-2025-010).

Theoretical analysis and microscopic modelling were supported by the Russian Science Foundation (project no. 24-19-00187 ; https://rscf.ru/project/24-19-00187 ).

\section*{Author contributions} 

%Example 
S.V.B. initiated and supervised the project.  A.S.S., A.G.Sh, I.A.G., and V.S.S. fabricated the devices;  R.T., D.S.K, B.V.F., and V.S.S. performed the current measurements;  R.T., D.S.K, S.V.B., I.A.G., N.V.K, A.S.S. and V.S.S. analysed the experimental data; I.A.G., R.T. and D.S.K performed spectra processing, 
S.V.B., N.V.K., M.Yu.K., and A.A.G. prepared the theoretical microscopic model; A.A.N. and I.I.S. made numerical calculations;  R.T., D.S.K, A.A.N, and S.V.B. prepared the initial draft; V.S.S., N.V.K., M.Yu.K. and A.A.G. reviewed and edited the manuscript with essential contributions from the other authors.

\bibliography{ref,MSR}

\end{document}